\documentclass[aps,pra,reprint,amssymb,amsmath,amsfonts,superscriptaddress,showpacs,floatfix]{revtex4-1} 
\usepackage{hyperref}
\hypersetup{colorlinks,citecolor=blue,filecolor=black,linkcolor=blue,urlcolor=black}

\usepackage{graphicx}

\usepackage{lmodern} 
\usepackage[caption=false]{subfig}
\usepackage{siunitx}
\usepackage{bm} 
\usepackage{color}
\usepackage[usenames,dvipsnames]{xcolor}
\usepackage{upgreek}
\usepackage{natbib}
\usepackage[normalem]{ulem}

\definecolor{ppurple}{RGB}{98,111,179}
\definecolor{porange}{RGB}{240,145,55}
\definecolor{pgreen}{RGB}{53,178,87}
\definecolor{pred}{RGB}{220,73,89}
\definecolor{pblue}{RGB}{7,174,227}

\newcommand*{\mathcolor}{}
\def\mathcolor#1#{\mathcoloraux{#1}}
\newcommand*{\mathcoloraux}[3]{%
  \protect\leavevmode
  \begingroup
    \color#1{#2}#3%
  \endgroup
}

\let\OldBlacksquare\blacksquare \renewcommand{\blacksquare}{{\mathcolor{black}{\OldBlacksquare}}}

\DeclareMathOperator\erf{erf}

\usepackage{textcomp}

\usepackage[nointegrals]{wasysym}
\usepackage{pifont}

\usepackage[geometry]{ifsym}
\let\OldBoxplus\boxplus \renewcommand{\boxplus}{{\mathcolor{black}{\OldBoxplus}}}
\newcommand{\grsquare}{\makebox(5,3)[lb]{\textcolor{black}{\FilledSmallSquare}}}

\usepackage{cleveref}

\usepackage{tikz}
\usetikzlibrary{shapes.geometric}

\crefname{figure}{Fig.}{Figs.}
\crefname{equation}{Eq.}{Eqs.}
\crefname{section}{Sec.}{Secs.}
\crefname{table}{Tab.}{Tabs.}

\definecolor{orange}{rgb}{1,0.5,0}

\begin{document}
\title{Experimental creation and characterization of random potential energy landscapes exploiting speckle patterns}
\author{J\"org Bewerunge}
\affiliation{Condensed Matter Physics Laboratory, Heinrich Heine University, 40225 D\"usseldorf, Germany}
\author{Stefan U. Egelhaaf}
\affiliation{Condensed Matter Physics Laboratory, Heinrich Heine University, 40225 D\"usseldorf, Germany}

\date{\today}


\begin{abstract}
The concept of potential energy landscapes is applied in many areas of science.
We experimentally realize a random potential energy landscape (rPEL) to which colloids are exposed.
This is achieved exploiting the interaction of matter with light.
The optical set-up is based on a special diffuser, which creates a top-hat beam containing a speckle pattern.
This is imposed on colloids.
The effect of the speckle pattern on the colloids can be described by a rPEL.
The speckle pattern as well as the rPEL are quantitatively characterized.
The distributions of both, intensity and potential energy values, can be approximated by Gamma distributions.
They can be tuned from exponential to approximately Gaussian with variable standard deviation, which determines the contrast of the speckles and the roughness of the rPEL.
Moreover, the characteristic length scales, e.g.~the speckle size, can be controlled.
By rotating the diffuser, furthermore, a flat potential can be created and hence only radiation pressure exerted on the particles.
\end{abstract}


\maketitle


\section{Introduction}
\label{sec:introduction}

A potential energy surface is a multi-dimensional surface that represents the potential energy of a system as a function of the coordinates and/or other parameters of its constituents, usually atoms, molecules or particles~\cite{Wales2004}.
Since its topographical features resemble a landscape with mountain ranges, valleys and passes, frequently it is referred to as a potential energy landscape (PEL), despite typically being multi-dimensional.
The PEL defines all the thermodynamic and kinetic properties of a system.
The evolution of a system can pictorially be described by the motion of a point on the PEL.

The concept of a PEL is successfully used in many fields of science to determine the properties and behavior of systems ranging from small to polymeric (bio)molecules and from atomic clusters to biological cells~\cite{Wales2004}.
They are used to describe, e.g., the particle dynamics in dense and crowded systems~\cite{Heuer2008,Debenedetti2001,Angell1995,Stillinger1995}, on surfaces~\cite{Jardine2004,Cordoba2014,Hsieh2014}, between magnetic domains~\cite{Tierno2010}, and in inhomogeneous materials~\cite{Chen2000,Weiss2004,Hofling2013,Tolic-Norrelykke2004} as well as the effects of external potentials on the dynamics of ultracold atoms~\cite{White2009,Robert2010}, quantum gases~\cite{Bouyer2010}, Bose-Einstein condensates~\cite{Lye2005,Fort2005,Clement2005,Clement2006}, and their applications to atom cooling and trapping~\cite{Horak1998}, and also include the investigation of the minimum energy conformations of molecules~\cite{Wales2004}, and the folding and association of proteins and DNA~\cite{Baldwin1994,Dill1997,Durbin1996,Frauenfelder1991,Janovjak2007}. 

Here we experimentally create a PEL to which colloidal particles are exposed and which changes, e.g., their arrangement and dynamics~\cite{Bouchaud1990, Dean2007, Sengupta2005, Isichenko1992, Goychuk2014, Banerjee2014, Wales2004, Zwanzig1988}.
As a model system, it can help to improve our understanding of the underlying principles governing the behavior in PELs and being common to different systems.

A PEL can experimentally be realized by exploiting the interaction of light with matter~\cite{Ashkin1986,Ashkin1992}.
We focus on large colloidal particles with a refractive index larger than the one of the dispersing liquid.
Their interaction with light is usually described by two forces ~\cite{Ashkin1986,Ashkin1992}: a scattering force or `radiation pressure', which pushes the particles along the beam, and a gradient force, which pulls particles towards regions of high intensity.
A classical application of this effect is optical tweezers which are used to trap and manipulate individual colloidal particles or groups of particles~\cite{Ashkin1986,Ashkin1992,Grier2003,Dholakia2008}.
Rather than tightly focused beams, extended light fields can be used to create a PEL \cite{Evers2013b}.
Light fields of almost any shape have been generated using spatial light modulators~\cite{Hanes2009,Hanes2012a,Hanes2012b,Hanes2013,Evers2013a,Evers2013b} or acousto-optic deflectors~\cite{Bowman2013,Neuman2004,Juniper2012}, while crossed laser beams~\cite{Ackerson1987,Jenkins2008b,Dalle-Ferrier2011} and other arrangements~\cite{Bechinger2001,Mikhael2008,Mikhael2010} have been used to create specific light fields.

Randomly-modulated intensity patterns, so-called laser speckles~\cite{Dainty1976,Goodman2007}, can be used to create a random potential energy landscape (rPEL).
The landscape can be rationalized as a superposition of many independent randomly-distributed optical traps.
They have been realized using various approaches:
holographic methods to produce one~\cite{Hanes2009,Hanes2012a,Hanes2012b, Hanes2013} and two-dimensional~\cite{Evers2013a,Mosk2012} patterns, optical fibers for two-dimensional patterns~\cite{Volpe2014} as well as diffusers for one~\cite{Boiron1999,Fort2005,Clement2005,Clement2006}, two~\cite{Horak1998,Lye2005,Brugger2015} and three-dimensional~\cite{White2009,Shvedov2010,Kondov2011,Douglass2012} patterns.

We use a special diffuser~\cite{Sales2003,Morris2007,Sales2012,Dickey2014} to create a random light field, that is a fully developed speckle pattern.
Due to the light-matter interactions, a colloidal particle exposed to the speckle pattern will experience a rPEL whose local value depends on the light intensity `detected' by the particle.
Since the particles are not point-like, the local potential value depends on the intensity distribution over the whole particle volume~\cite{Chowdhury1991,Loudiyi1992,Pelton2004,Jenkins2008b}.
We describe the interaction of a colloidal particle with the speckle pattern analogous to a detector that records the speckle intensity over a finite area.
This allows us to quantitatively characterize the statistics of the rPEL.
As will be shown, the distribution of energy values can be described by a Gamma distribution, and thus ranges from an exponential to an approximately Gaussian distribution, and the correlation length is set by the particle and speckle sizes.
The shape and width of the distribution and the correlation length hence can be tuned in a broad range.
The obtained rPEL can be applied to study the spatial arrangement and dynamics of colloidal particles in an external potential~\cite{Bouchaud1990,Dean2007,Isichenko1992,Goychuk2014,Banerjee2014,Zwanzig1988, Hanes2012a,Hanes2012b, Hanes2013,Evers2013a,Evers2013b}.
The chosen diffuser allows the creation of a large light field and thus the simultaneous investigation of many particles, which typically results in excellent statistics.
Furthermore, its small and compact design simplifies its alignment, movement and rotation.

\section{Creation of Speckle Patterns}
\label{sec:creation}

\begin{figure} 
\includegraphics[width=1.0\linewidth]{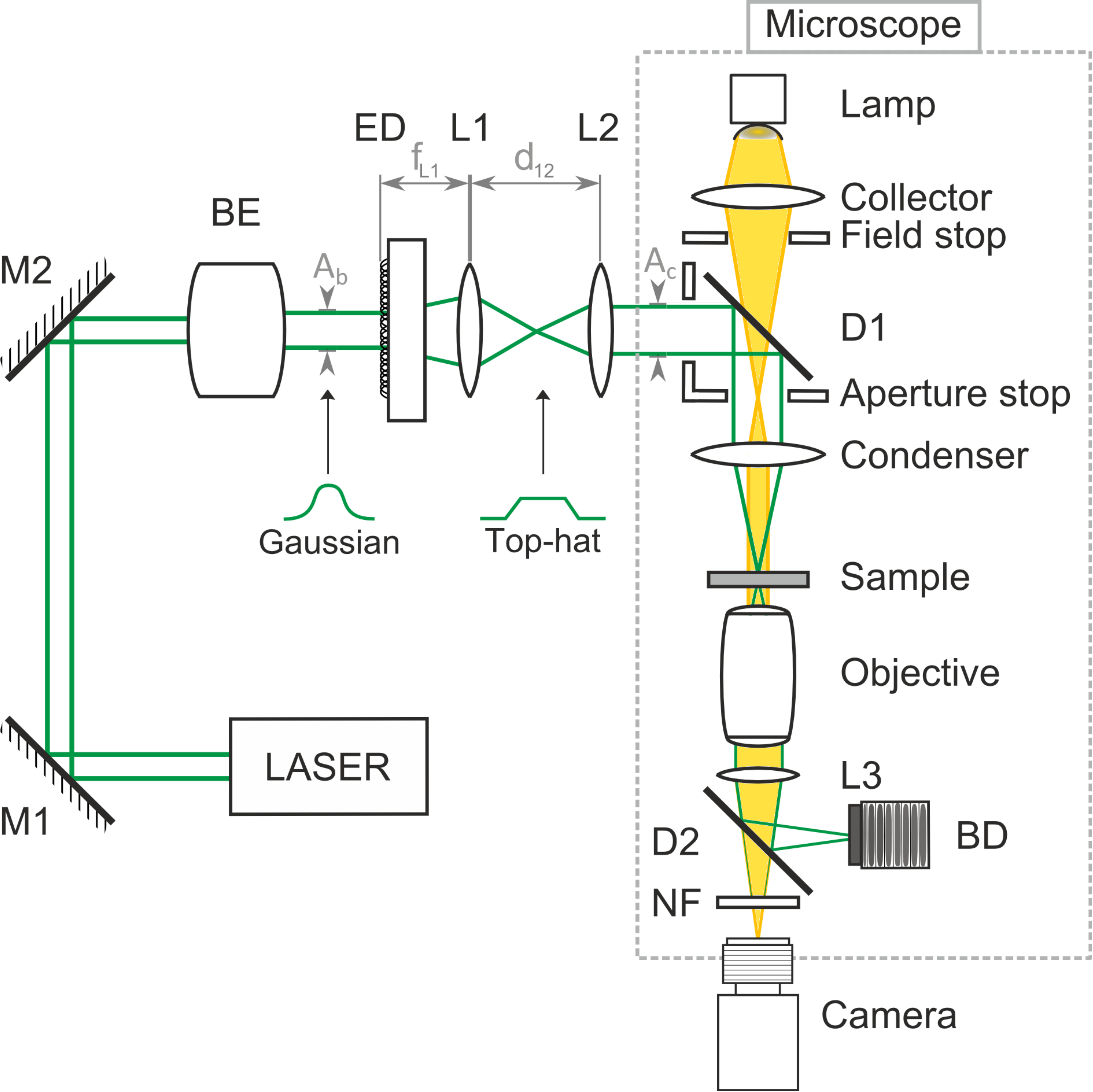}
\caption[Schematic diagram of the optical set-up]
{(Color online) Schematic representation of the set-up used to create a speckle pattern, to which colloidal particles are exposed and simultaneously imaged with an optical microscope.
The central optical element is a special diffuser (ED).
It is illuminated by a parallel Gaussian beam and creates a top-hat beam including a speckle pattern, which is steered to the sample plane of an inverted microscope.
See text for details.
\label{fig:set_up}}
\end{figure}

The set-up (Fig.~\ref{fig:set_up}) allows one to create a top-hat beam with a speckle pattern.
Thus, there are intensity fluctuations on a small length scale, i.e.~about the size of the colloidal particles.
At the same time, the top-hat beam implies a constant intensity on a larger length scale, at least the field of view.
This light field is used to impose a rPEL, without any underlying long-range variations, on colloidal particles.
The particles are constrained to a quasi two-dimensional plane and can simultaneously be observed with an optical microscope.

\subsection{Diffuser}
\label{sec:specklegen}

\begin{figure}  
\includegraphics[width=0.9\linewidth]{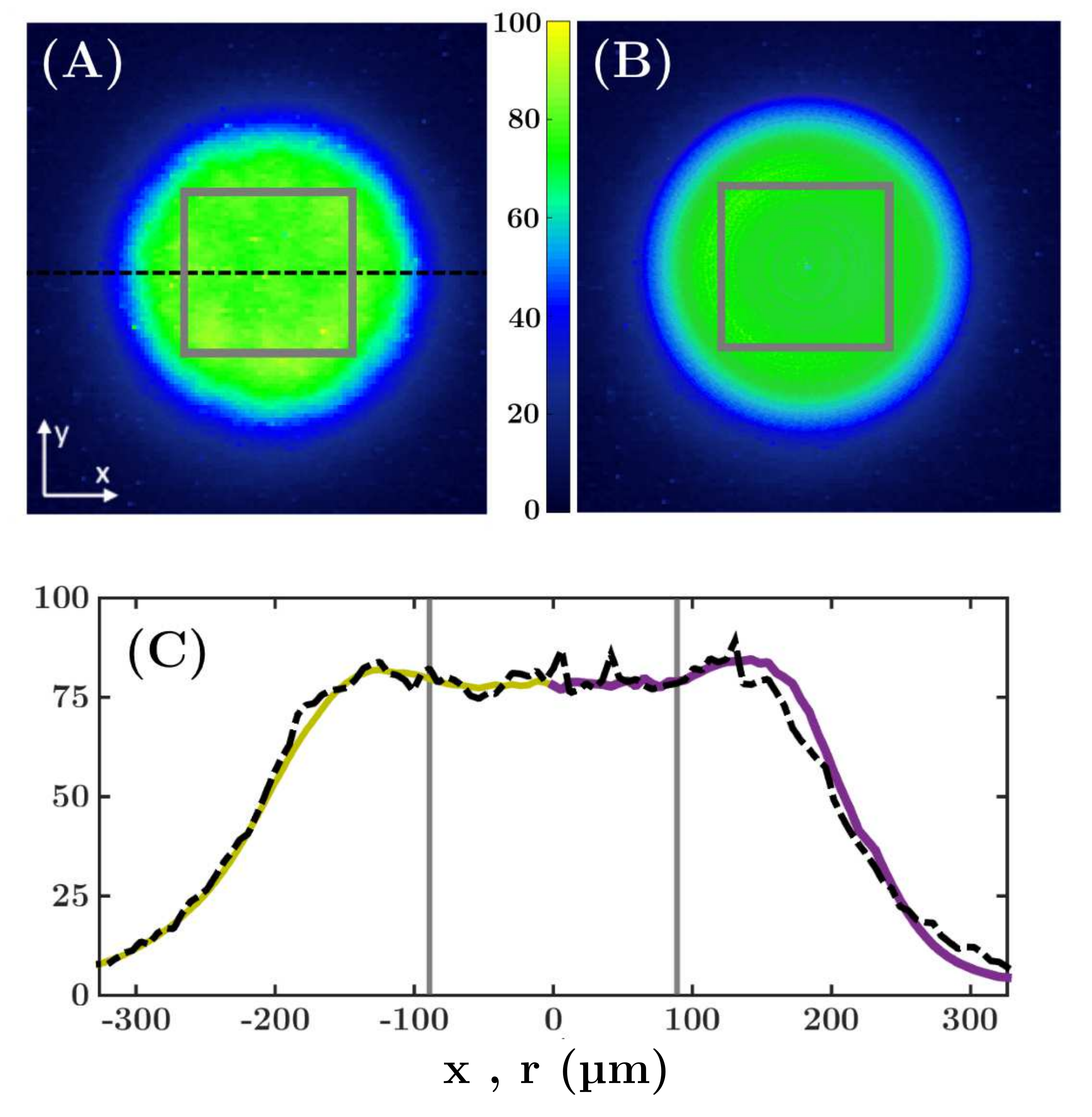}
\caption[Normalized intensity pattern (A), time-avergaed intensity pattern (B) and two-dimensional profile (C)]
{(Color online) (A) Individual and (B) averaged intensity patterns in the sample plane.
The average is taken over $120$ images with an individual exposure time of $10$~ms obtained with a rate of $10$~fps while the diffuser is rotated with a constant angular velocity of $20\,^\circ/$s.
The intensities in arbitrary units are represented by colors (as indicated).
(C) Corresponding intensity profiles along the horizontal dashed line in (A) (dashed line) and azimuthal average of the pattern in (A) (yellow line on the left) and in (B) (purple line on the right).
Experimental conditions BE~5$\times$ (Tab.~\ref{tab:speckles}).
Measurements are performed with a beam profiler (Coherent LaserCam HR).
Grey rectangles in (A) and (B) and grey lines in (C) indicate a field of view of $179 \times 179~\upmu$m$^2$.
}
\label{fig:profile}
\end{figure}

\begin{figure} 
\includegraphics[width=0.75\linewidth]{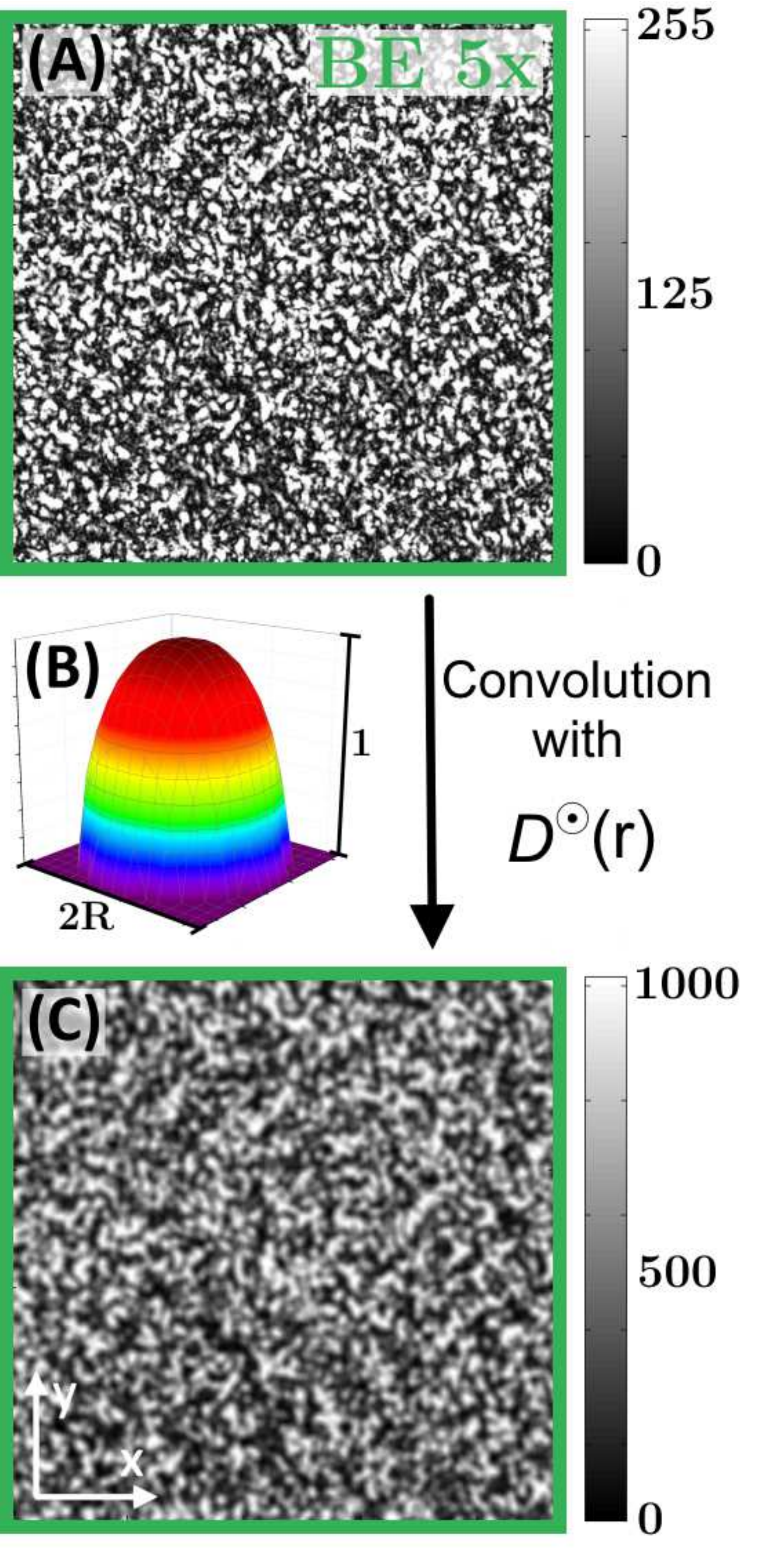}
\caption[Speckle image, detector function and approximated potential]
{(Color online) (A) Speckle pattern filling a field of view of $108 \times 108~\upmu$m$^2$ corresponding to 480~$\times$~480 pixels with intensities represented as grey levels (as indicated).
Experimental conditions BE~5$\times$ (\cref{tab:speckles}).
(B) Weight function $D^{\odot}(\mathbf{r})$ (Eq.~\ref{eq:ID}) representing the volume of a spherical colloidal particle with radius $R=1.4\;\upmu$m.
(C) Random potential energy landscape (rPEL) experienced by the particle in the speckle pattern shown in (A) and calculated by convolving the intensity in (A) with $D^{\odot}(\mathbf{r})$.
The values of the potential in arbitrary units are represented as grey levels (as indicated).
\label{fig:IplusUplusConv}}
\end{figure}

The central optical element of the set-up is a special diffuser (RPC Photonics, Engineered Diffuser\texttrademark\, EDC-1-A-1r, diameter $25.4$~mm)~\cite{Sales2003, Sales2012, Dickey2014}.
It is a laser-written, randomly-arranged array of microlenses that vary in radius of curvature and size and cover on average an area $A_{\text{l}} \approx 2000~\upmu$m$^2$.
When illuminated with an expanded Gaussian laser beam, individual wavefronts originate from each microlens whose characteristics are designed such that a macroscopically uniform intensity pattern with a small divergence is produced, reflecting a top-hat intensity distribution (\cref{fig:profile})~\cite{Korotkova2013,Dickey2014}.
Nonetheless, the random distribution as well as the individual variations of the microlenses and the interference of the corresponding wavefronts lead to microscopic intensity variations, i.e.~laser speckles (Figs.~\ref{fig:profile}A, \ref{fig:IplusUplusConv}A).
The speckle pattern consists of three-dimensional cylindrical high-intensity regions~\cite{Li2012}.
Their orientation and position with respect to the beam axis determine the properties of the speckles in the two-dimensional sample plane~\cite{Li2011a,Li2011b}.
Thus the correct imaging of the modified beam into the sample plane of the microscope is important.
Moreover, the speckle size is controlled by the diameter of the illuminating laser beam, determining the number of illuminated microlenses.
Their number is chosen large enough to ensure a statistically fully developed speckle pattern~\cite{Dainty1976,Goodman2007}.
By changing the position of the beam on the diffuser, statistically equivalent but independent realizations of the speckle pattern can be created.

\subsection{Optical Set-up}
\label{sec:setup}

The speckle pattern strongly depends on the properties of the beam incident on the diffuser.
Fully developed speckles require the interference of many polarized monochromatic wavefronts with random phases and amplitudes and thus a large incident beam that illuminates many microlenses.
Furthermore, the optics used to image the modified beam into the sample plane, especially their apertures, have to be designed carefully and, for imaging the speckle pattern, also the detector and its pixel size have to be considered.

A solid-state laser (Laser Quantum, Opus 532, wavelength $\lambda = 532$~nm, maximum intensity $P_{\mathrm{L,max}} = 2.6$~W) provides a monochromatic linearly-polarized Gaussian beam which is slightly elliptical with an axial ratio of $1.12$.
The laser beam is steered by two mirrors (M1, M2, \cref{fig:set_up}) to a beam expander (BE, Sill Optics, S6EXZ5076/121) with variable magnification ($1-8\times$) and divergence correction.
Using the beam expander, the area $A_{\text{b}}$ of the Gaussian beam hitting the diffuser can be controlled.
The diffuser is mounted in a motorized rotation stage (Newport, PR50CC).

The beam leaving the diffuser is divergent (about $1^{\circ}$) and hence collimated by two lenses (L1, L2), where the first lens (L1, Edmund Optics, 1"DCX75, focal length $f_{\mathrm{L1}}=7.5$~cm) is placed a distance $f_{\mathrm{L1}}$ behind the diffuser followed by the second lens (L2, Thorlabs, 2"PCX75, $f_{\mathrm{L2}}=7.5$~cm) in a distance $d_{12}=16$~cm.
This leads to a collimated beam with area $A_{\text{c}}$ (about $1\:$cm$^2$ for a 5$\times$ magnification of the beam expander, i.e.~BE~5$\times$, Tab.~\ref{tab:speckles}) on the aperture stop, after the beam has been introduced into the light path of the inverted microscope by a dichroic mirror (D1, Edmund Optics, NT69-901).
The condenser (Nikon, TI-C-LWD) then focuses the beam in the sample plane (Figs.~\ref{fig:profile}A, \ref{fig:IplusUplusConv}A).
The lenses (L1, L2) together with the condenser form a telecentric illumination system which collimates the beam and focuses it in the sample plane.

The laser beam is removed from the light path of the microscope by a dichroic mirror (D2, Edmund Optics, NT69-901) which deflects the beam into a beam dump (BD).
Furthermore, a notch filter (NF, Edmund Optics NT67-119, optical density OD4 at $\lambda=532$~nm) is introduced in front of the camera.

The colloidal particles are observed using an inverted microscope (Nikon, Eclipse Ti-U) with usually a $20\times$ objective
(Nikon, CFI S Plan Fluor ELWD, N.A.~$0.45$) and an optional additional magnification of $1.5 \times$ resulting in a field of view of $431 \times 345~\upmu$m$^2$ and $288 \times 230~\upmu$m$^2$, respectively.
The images are recorded using an 8-bit CMOS camera (PixeLINK, PL-B741F with $1280 \times 1024$ pixels, if not stated otherwise).

To image the speckle pattern at low laser intensities $P_{\mathrm{L}} \approx 1$~mW, the dichroic mirror (D2) and notch filter (NF) are removed.
When examining the speckle pattern, a very dilute sample (less than five particles in the field of view) is used.
The sedimented particles help to focus on the sample plane and hence record the relevant plane of the speckle pattern.
The presence of a sample also leaves the light path unchanged.
This ensures that the recorded speckle pattern represents the intensity distribution to which the particles are exposed.


\section{Characteristics of Speckle Patterns}
\label{sec:specklechar}

If wavefronts of the same wavelength but with random phases and amplitudes, as those created by the microlenses, interfere, speckle patterns occur.
Speckles are characterized by intensity fluctuations on a small length scale but a uniform intensity on a larger length scale.
The statistics of the intensity fluctuations, such as the intensity distribution and spatial correlation, have been investigated in the context of coherent light reflected from rough surfaces or transmitted through diffusers~\cite{Dainty1976, Dainty1984, Goodman2007}.
The same statistics are expected for the speckle pattern created by the present diffuser~\cite{Sales2003,Sales2012,Dickey2014}. Thus below we follow~\cite{Dainty1976, Dainty1984, Goodman2007}.

\subsection{Ideal Speckles}

The interference of many monochromatic and linearly polarized wavefronts with random phasors results in a fully developed speckle pattern.
In this case, the intensity distribution of the speckle pattern follows an exponential distribution
\begin{equation}
p(I) = \frac{1}{\langle I \rangle} \, \exp{\left(- \frac{I}{\langle I \rangle}\right)}
\label{eq:pI}
\end{equation}
with the mean intensity $\langle I \rangle$ and standard deviation $\sigma= ( \left \langle I^2 \right\rangle - \langle I \rangle^2 )^{1/2}=\langle I \rangle$.

The normalized standard deviation represents the contrast of the speckle pattern,
\begin{equation}
c = \frac{\sigma}{\langle I \rangle} = \frac{\sqrt{\langle I^2 \rangle - \langle I \rangle^2}}{\langle I \rangle} \ \text{.}
\label{eq:contrast}
\end{equation}
The contrast $c$ quantifies the magnitude of the intensity fluctuations.
For an exponential distribution, i.e.~a fully developed speckle pattern, it reaches its maximum value $c=1$.

The spatial structure of the speckle pattern is characterized by the normalized spatial autocorrelation function of the intensity~\cite{Goodman2007,Dainty1976,Dainty1984,Li2012}
\begin{equation}
C_{\mathrm{I}}(\Delta {\mathbf{r}}) = \frac{\left \langle I(\mathbf{r})I(\mathbf{r} + \Delta {\mathbf{r}})\right \rangle}{\left \langle I(\mathbf{r})\right \rangle^2} - 1 \ \text{,}
\label{eq:intautocorr}
\end{equation}
where $I(\mathbf{r})$ is the intensity at position $\mathbf{r}$ and $\langle \; \rangle$ can represent both, an ensemble or a spatial average.
A spatially infinite pattern without long-range correlations is self-averaging~\cite{Lifshits1988} and hence the spatial and ensemble averages coincide.
To a good approximation this also holds for (finite) experimental speckle patterns, similar to the ones considered here~\cite{Clement2006}.
The extent of $C_{\mathrm{I}}(\Delta {\mathbf{r}})$ provides a measure for the correlation area of the speckle pattern, that is the characteristic speckle area
\begin{equation}
A_{\mathrm{S}} = \iint\limits_{-\infty}^{\hspace{10pt}\infty}  \! C_{\mathrm{I}}(\Delta {\mathbf{r}}) \, \mathrm{d}^2\Delta {\mathbf{r}} \ \text{.}
\label{eq:AS}
\end{equation}

\subsection{Integrated Speckles}

In an experimental situation, the optical elements and especially their apertures as well as the finite detector size have to be considered~\cite{Dainty1976,Dainty1984,Goodman2007,Skipetrov2010}.
The finite detector size can be taken into account through the weight function $D(\mathbf{r})$, which represents the spatial sensitivity of the detector.
Accordingly, the effective detector area $A_{\mathrm{D}}$ can be calculated as
\begin{equation}
A_{\mathrm{D}} = \iint\limits_{-\infty}^{\hspace{10pt}\infty} \! D(\mathbf{r}) \, \mathrm{d}^2\mathbf{r} \ \text{.}
\label{eq:AD}
\end{equation}

In the following also the (deterministic) autocorrelation function of the weight function $D(\mathbf{r})$ is required, which is given by
\begin{equation}
C_{\mathrm{D}}(\Delta \mathbf{r}) = \frac{1}{A_{\mathrm{D}}} \iint\limits_{-\infty}^{\hspace{10pt}\infty} \! D(\mathbf{r})D(\mathbf{r}{-}\Delta \mathbf{r}) \, \mathrm{d}^2\mathbf{r} \ \text{.}
\label{eq:CD}
\end{equation}
Based on $C_{\mathrm{D}}(\Delta \mathbf{r})$, the effective measurement area is defined as
\begin{equation}
A_{\text{m}} = \frac{A_{\mathrm{D}}}{C_{\mathrm{D}}(\mathbf{0})}
   = \frac{A_{\mathrm{D}}^2}{\iint\limits_{-\infty}^{\hspace{10pt}\infty} \! D^2(\mathbf{r}) \, \mathrm{d^2}\mathbf{r}}\ \text{.}
\label{eq:Am}
\end{equation}

A detector centered at position $\mathbf{r}$ registers an intensity $I_{\mathrm{D}}(\mathbf{r})$ that is the integrated intensity taking the weight function $D(\mathbf{r})$ into account, i.e.~\cite{Dainty1976,Skipetrov2010}
\begin{equation}
I_{\mathrm{D}}(\mathbf{r}) = \frac{1}{A_{\mathrm{D}}} \iint\limits_{-\infty}^{\hspace{10pt}\infty} \! D(\Delta \mathbf{r})I(\mathbf{r}{+}\Delta \mathbf{r}) \, \mathrm{d}^2\Delta \mathbf{r}\ \text{.}
\label{eq:ID}
\end{equation}
The intensity distribution for a finite detector is described to a good approximation by a Gamma distribution
\begin{equation}
p(I_{\text{D}}) = \frac{1}{\Gamma(M)} \left(\frac{M}{\langle I_{\text{D}} \rangle}\right)^{M} I_{\text{D}}^{M-1} \,  \exp \left(-\frac{M}{\langle I_{\text{D}} \rangle}I_{\text{D}}\right)\ \text{,}
\label{eq:PIP}
\end{equation}
where $\Gamma$ is the Gamma function and the mean of the detected intensity is identical with the mean of the ideal speckle pattern, i.e.~$\langle I_{\text{D}} \rangle =  \langle I \rangle$, and the normalized standard deviation or contrast is $c_\text{D} = 1/M^{1/2}$, if noise and correlations between neighboring pixels are absent.
The parameter $M$ is given by
\begin{equation}
M = \left ( \frac{1}{A_{\mathrm{D}}} \iint\limits_{-\infty}^{\hspace{10pt}\infty} \! C_{\mathrm{I}}(\Delta \mathbf{r}) C_{\mathrm{D}}(\Delta \mathbf{r})  \, \mathrm{d^2}\Delta\mathbf{r} \right )^{-1}
 \ \text{,}
\label{eq:M}
\end{equation}
which depends on the spatial characteristics of the speckle pattern and detector, i.e.~the correlation functions of the intensity $C_{\mathrm{I}}(\Delta \mathbf{r})$ (Eq.~\ref{eq:intautocorr}) and weight function $C_{\mathrm{D}}(\Delta \mathbf{r})$ (Eq.~\ref{eq:CD}), respectively.

If the effective measurement area $A_{\text{m}}$ is large compared to the speckle area $A_{\text{S}}$, i.e.~$A_{\text{m}} \gg A_{\text{S}}$, many speckles contribute to the detected intensity $I_{D}(\mathbf{r})$.
Then $M$ represents the (large) number of detected speckles, $M \approx A_{\text{m}} / A_{\text{S}} \gg 1$~\cite{Dainty1976,Goodman2007,Skipetrov2010}, and $p(I_{\text{D}})$ approaches a Gaussian distribution with mean $\langle I \rangle$ and normalized standard deviation $c$.
In the opposite limit of a very small effective measurement area, $A_{\text{m}} \ll A_{\text{S}}$, only one speckle is detected.
Thus $M \to 1$ and $p(I_{\text{D}})$ approaches the exponential distribution (Eq.~\ref{eq:pI}).
In this case, neighboring detectors might no longer be independent.
If, however, the effective measurement area and speckle area are similar ($A_{\text{m}} \approx A_{\text{S}}$), M can only be (numerically) calculated if $C_{\mathrm{I}}(\Delta \mathbf{r})$ and $D(\mathbf{r})$ are known.
Due to the complex effects of the optical components on the speckle pattern, this often is not the case and approximations must be used.

In the following we apply these relationships for different detectors and thus $D({\bf r})$ (the different cases are indicated by superscripts); circular ($D^{\text{\textcolor{black}{\FilledSmallCircle}}}({\bf r})$) and square~($D^{{\text{\grsquare}}}(\mathbf{r})$) detector pixels, which are also subjected to smoothing ($D^{\text{\textcolor{black}{\FilledSquareShadowC}}}({\bf r})$) and binning ($D^{\mathcolor{black}{\boxplus}}({\bf r})$), as well as spherical ($D^{\odot}({\bf r})$) and cubic ($D^{\boxdot}({\bf r})$) particles acting as `detectors'.
%
%

\subsubsection{Detector Pixel}
\label{sec:pixel}

In our experiments, the speckle patterns are detected by uniform square pixels.
Their weight function is
\begin{equation}
D^{{\text{\grsquare}}}(\mathbf{r}) =  \left \{
   \begin{array}{c@{\quad \quad}l}
    1 & {\text{inside the pixel}} \\
    0 & {\text{outside the pixel}}
    \end{array} \right. 
\label{eq:D_pixel}
\end{equation}
and hence $A_{\text{m}}^{{\text{\grsquare}}} = A_{\mathrm{D}}^{{\text{\grsquare}}}$ are equal to the pixel area.
Based on this weight function, $C_{\text{D}}^{\text{\grsquare}}(\Delta \mathbf{r})$ (Eq.~\ref{eq:CD}) and $I_{\text{D}}^{\text{\grsquare}}(\mathbf{r})$ (Eq.~\ref{eq:ID}) can be calculated.
Furthermore, it is expected that $p(I_{\text{D}}^{\text{\grsquare}})$ can be approximated by a Gamma distribution (Eq.~\ref{eq:PIP}).
However, to calculate the parameter $M^{\text{\grsquare}}$ (Eq.~\ref{eq:M}), also $C_{\text{I}}^{\text{\grsquare}}(\Delta \mathbf{r})$ (Eq.~\ref{eq:intautocorr}) is required.

If the top-hat beam is approximated by a Gaussian beam, the corresponding result for a Gaussian beam detected by uniform square pixels~\cite{Skipetrov2010,Goodman2007},
\begin{equation}
\label{eq:Gaussian}
C_{\text{I}}^{{\text{\grsquare}}}(\Delta \mathbf{r}) = \exp{\left ( - \frac{\pi \Delta \mathbf{r}^2}{A_{\text{s}}} \right )} \ \text{, \;\;}
\end{equation}
can be used.
Then, $M^{{\text{\grsquare}}}$ is given by
\begin{align}
M^{{\text{\grsquare}}} = &\Bigg[\sqrt{\frac{A_{\text{S}}}{A_{\text{m}}^{{\text{\grsquare}}}}} \erf\left(\sqrt{\frac{\pi A_{\text{m}}^{{\text{\grsquare}}}}{A_{\text{S}}}}\right)\nonumber \\
&~-\left(\frac{A_{\text{S}}}{\pi A_{\text{m}}^{{\text{\grsquare}}}}\right)\left\{1-\exp\left(-\frac{\pi A_{\text{m}}^{{\text{\grsquare}}}}{A_{\text{S}}}\right)\right\}\Bigg]^{-2} \; \text{.} 
\label{eq:M2}
\end{align}
For a Gaussian beam detected by uniform circular pixels
\begin{align}
M^{\text{\textcolor{black}{\FilledSmallCircle}}} = &\frac{A_{\text{m}}^{\text{\textcolor{black}{\FilledSmallCircle}}}}{A_{\text{S}}} \Bigg[1 - \exp{\left (-\frac{2A_{\text{m}}^{\text{\textcolor{black}{\FilledSmallCircle}}}}{A_{\text{S}}} \right )}\nonumber \\
&~\times \left \{ I_0 \left ( \frac{2A_{\text{m}}^{\text{\textcolor{black}{\FilledSmallCircle}}}}{A_{\text{S}}} \right ) + I_1 \left ( \frac{2A_{\text{m}}^{\text{\textcolor{black}{\FilledSmallCircle}}}}{A_{\text{S}}} \right ) \right \} \Bigg]^{-1}
\; \text{,}
\label{eq:Mcirc}
\end{align}
where $I_0$ and $I_1$ are modified Bessel functions of the first kind and orders zero and one, respectively.
Further geometries have been considered~\cite{Dainty1976,Clement2006,Skipetrov2010,Li2012}, but are less appropriate for the present situation.

To check the suitability of the above equations for our experimental situation, in particular the approximation of the top-hat beam by a Gaussian beam, these relations will be compared to our experimental results in~\cref{sec:speckles}.

\subsubsection{Colloidal Particle}
\label{sec:potential}

Colloidal particles are susceptible to electromagnetic radiation if their refractive index is different from the one of the suspending liquid~\cite{Ashkin1986, Ashkin1992}.
Since the particles are not point-like, their response depends on the intensity integrated over their volume~\cite{Chowdhury1991,Loudiyi1992,Pelton2004,Jenkins2008b}.
This is analogous to the extended detector described above, except that the particle's susceptibility (or polarizability) rather than the detector efficiency is relevant.
It is proportional to the particle volume traversed by the beam.
Since the speckles are oriented in beam direction and their extension in beam direction is much larger than in the sample plane~\cite{Goodman2007,Li2012}, the projection of the particle volume in beam direction is considered.
The (projected) particle volume is taken into account through the weight function $D^{\odot}(r)$.
For a homogeneous spherical particle the normalized weight function is
\begin{equation}
D^{\odot}(r) 
  = \begin{cases}\frac{1}{R}\sqrt{{R}^{2}-r^{2}} & \mbox{if } r\leq R\\ 0 & \mbox{if } r>R\end{cases} \, \text{,}
\label{eq:D_particle}
\end{equation}
and shown in \cref{fig:IplusUplusConv}B.
To obtain its absolute value, material specific parameters describing the light--particle interaction have to be considered~\cite{Ashkin1986, Ashkin1992,Pelton2004} and summarized in a ($r$-independent) prefactor.
Independent of this constant prefactor, the effective measurement area (Eq.~\ref{eq:Am}), or rather effective particle area, becomes
\begin{equation}
\label{eq:Aeffcirc}
A_{\text{m}}^{\odot} = \frac{8\pi}{9} R^2 \, \text{.}
\end{equation}
The (deterministic) autocorrelation function of the weight function $D^{\odot}(r)$, that is $C_{\text{D}}^{\odot}(\Delta \mathbf{r})$ (Eq.~\ref{eq:CD}), can only be determined numerically~\cite{Pelton2004}.
Finally, taking into account the particle volume through $D^{\odot}(r)$, the integrated intensity $I_{\text{D}}^{\odot}({\mathbf{r}})$ can be calculated (Eq.~\ref{eq:ID})~\cite{Chowdhury1991,Loudiyi1992, Pelton2004}.

Exploiting the analogy between a colloidal particle and a detector, we expect that the intensity distribution as experienced by the particle, i.e.~the rPEL, can be approximated by a Gamma distribution, similar to Eqs.~\ref{eq:PIP} and~\ref{eq:M}, but its parameter $M$ has to be determined.
This analogy is explored and experimentally tested in \cref{sec:rPEL}.


\section{Results and Discussion}
\label{sec:results}


\subsection{Speckle Pattern}
\label{sec:speckles}

Different speckle patterns are created by changing the size of the beam that illuminates the diffuser using the variable beam expander.
Magnifications between $3 \times$ and $7 \times$ are possible yielding beam areas on the diffuser $0.3$~cm$^2$~$\lesssim A_{\text{b}} \lesssim 1.7$~cm$^2$ (\cref{tab:speckles}).
Stationary speckle patterns as well as time-varying speckle patterns, created by rotating the diffuser, are investigated and the data compared to the relations presented above (\cref{sec:pixel}) to test their applicability to the present experimental situation.

\begin{table*} 
\caption{Experimental conditions with different magnifications of the beam expander, where the nominal magnifications serve as labels.
Parameters characterizing the speckle patterns, namely the ratio of the beam area to the mean microlens area, $A_{\text{b}}/A_{\text{l}}$ where $A_{\text{l}} \approx 2000~\upmu\text{m}^2$, speckle contrast $c$ (Eq.~\ref{eq:contrast}), and speckle area $A_{\text{S}}^{\text{\textcolor{black}{\FilledSmallSquare}}}$ (Eq.~\ref{eq:AS}). Parameters characterizing the speckle patterns convolved with the weight function $D^\odot(r)$ of a particle with radius $R=1.4~\upmu{\text{m}}$ and thus $A_{\text{m}}^{\odot}=5.5~{\upmu\text{m}}^2$, i.e.~parameters characterizing the rPEL, namely the ratio of the effective particle area to the speckle size, $A_{\text{m}}^{\odot}/A_{\text{S}}^{\text{\textcolor{black}{\FilledSmallSquare}}}$, the parameter M, the correlation area $A_{\text{S}}^{U}$, the effective correlation area of the weight function $A_{\text{S}}^{\odot} = A_{\text{S}}^{U} - A_{\text{S}}^{\text{\textcolor{black}{\FilledSmallSquare}}}$. Furthermore, the corresponding symbols used in the figures are indicated.
}
\label{tab:speckles}
\includegraphics[width=1.0\linewidth]{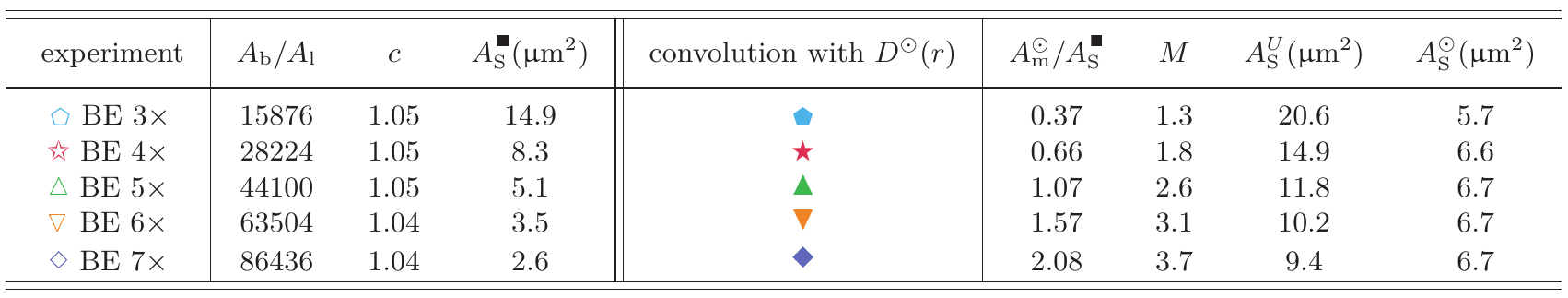}
\end{table*}

\subsubsection{Stationary Speckle Pattern}

\begin{figure} 
\includegraphics[width=1.0\linewidth]{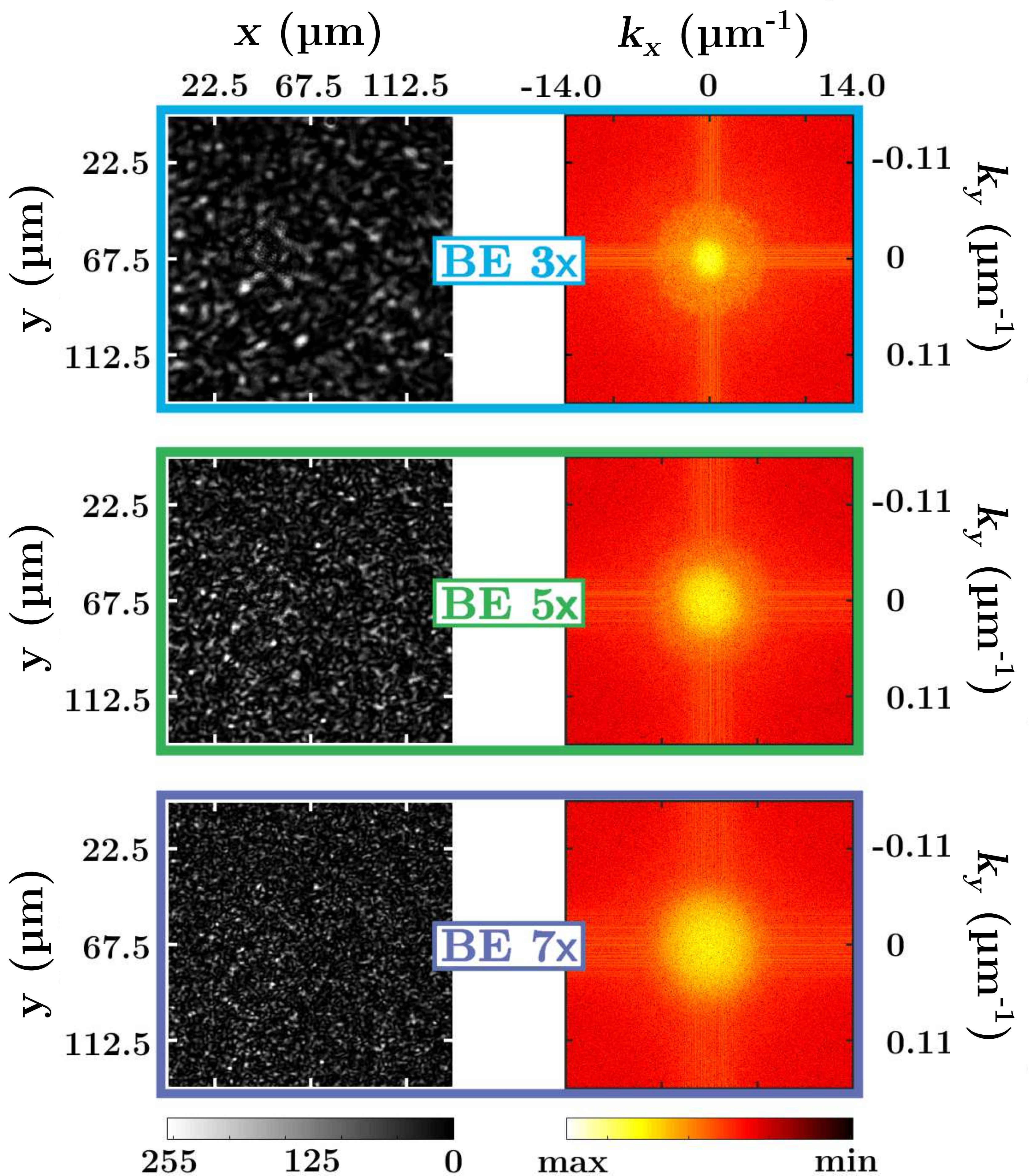}
\caption[$I(x,y)$ and PSD for different speckle areas]
{(Color online) (left) Speckle patterns $I_{\text{D}}^{\text{\grsquare}}({\mathbf{r}})$ created with different beam areas $A_{\text{b}}$ due to different magnifications of the beam expander (as indicated, Tab.~\ref{tab:speckles}).
Intensities are represented as grey levels (scale at bottom).
(right) Corresponding power spectral densities with their values represented by colors (logarithmic scale in arbitrary units at bottom).
For clarity the lowest frequencies are shifted to the center.
The spurious high values in $x$ and $y$ direction through the origin are caused by boundary effects in the Fourier transform.
\label{fig:IvsPSD}
}
\end{figure}

\begin{figure} 
\includegraphics[width=0.9\linewidth]{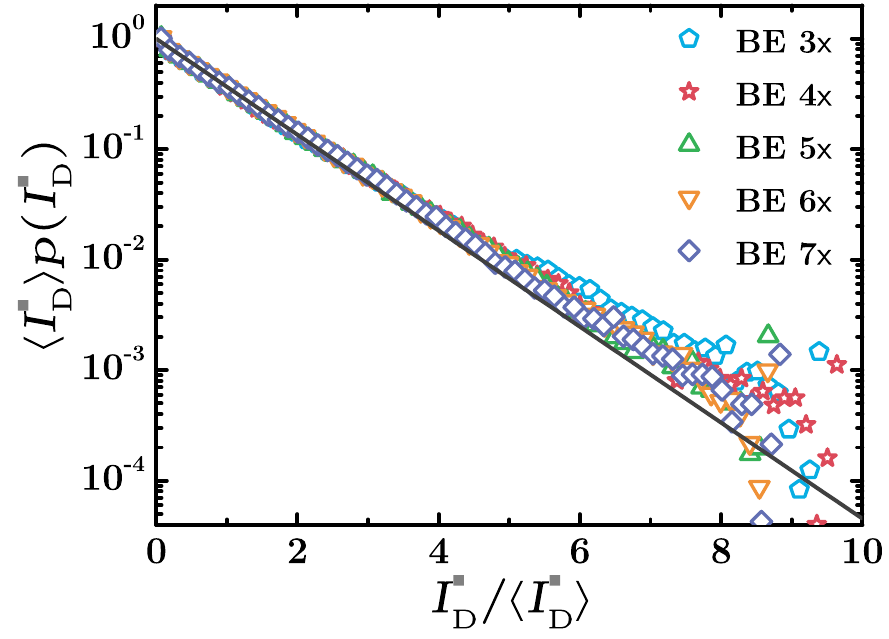}
\caption[Normalized probability distribution of the intensity $p(I)$]
{(Color online) Normalized intensity distributions $\langle I_{\text{D}}^{\text{\grsquare}} \rangle \; p(I_{\text{D}}^{\text{\grsquare}})$ as observed in experiments with different beam areas $A_{\text{b}}$ due to different magnifications of the beam expander (as indicated, \cref{tab:speckles}).
Each symbol is the average of four data points.
The line represents an exponential distribution (Eq.~\ref{eq:pI}).
}
\label{fig:PI}
\end{figure}

The observed intensities $I_{\text{D}}^{\text{\grsquare}}({\mathbf{r}})$ (Figs.~\ref{fig:IplusUplusConv}A, \ref{fig:IvsPSD}, left) resemble speckle patterns with their characteristic intensity fluctuations.
A qualitative inspection indicates a decreasing speckle size with increasing beam size.
The magnitude of the intensity fluctuations is quantified by the intensity distribution $p(I_{\text{D}}^{\text{\grsquare}})$ (\cref{fig:PI}) and the contrast $c$ (\cref{tab:speckles}).
The observed $p(I_{\text{D}}^{\text{\grsquare}})$ are well described by an exponential distribution (Eq.~\ref{eq:pI}), which suggests fully developed speckles.
This is consistent with the fact that all beam areas $A_{\text{b}}$ are much larger than the microlens area $A_{\text{l}}$ and hence many microlenses ($A_{\text{b}}/A_{\text{l}} > 10^4$) are illuminated and, in addition, the detector pixels are much smaller than the speckle area, i.e.~$A_{\text{m}}^{\text{\grsquare}} \ll A_{\text{S}}^{\text{\grsquare}}$.
Only small deviations from an exponential distribution are observed.
The smallest intensity occurs with a slightly larger probability.
This is attributed to the finite exposure time and sensitivity of the camera, which limit the minimum detectable intensity.
If, within the exposure time, too few photons are registered, the pixel will record zero intensity which thus occurs with a slightly too large probability.
Also the highest intensities are recorded slightly too frequently due to noise together with the limited dynamic range of the 8-bit camera given the large range of intensity values.
Still, the chosen exposure time and laser power provide the optimum compromise.

The normalized standard deviation of $p(I_{\text{D}}^{\text{\grsquare}})$ or contrast $c$ (Eq.~\ref{eq:contrast}) is found to be close to one (\cref{tab:speckles}), which is consistent with fully developed speckles.
However, the contrast is slightly larger than one.
This might be due to the flat-top instead of a Gaussian beam~\cite{Baykal2013} and additional noise, for example contributed by the camera~\cite{Song2013}.
The depolarization and scattering by the (very few) particles in the sample plane might also contribute.
The increase of $c$ with decreasing beam area $A_{\text{b}}$ is attributed to slight changes in the divergence of the beam that has not been corrected in this series.

\begin{figure} 
\includegraphics[width=0.9\linewidth]{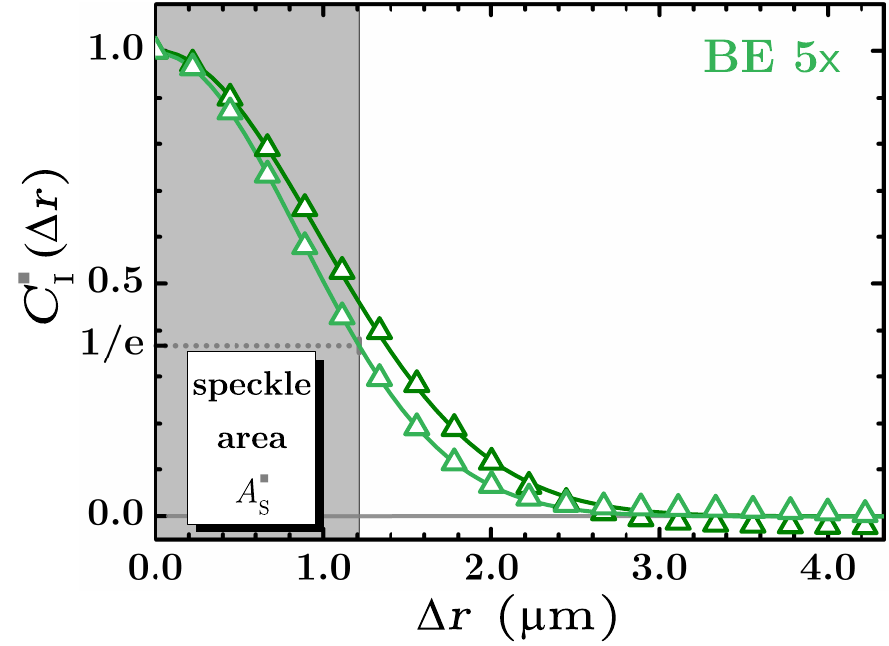}
\caption[Intensity autocorrelation $C_{\text{I}}$ as a function of $\mathbf{\Delta r}$]
{(Color online) Intensity correlation function $C_{\text{I}}^{\text{\grsquare}}$ as a function of $\Delta r$ in $x$ 
(light green triangle, left) and $y$ 
(dark green triangle, right) directions as observed in the experiment BE~5$\times$ (Tab.~\ref{tab:speckles}).
Predictions for a Gaussian beam detected by square pixels $C_{\text{I}}^{{\text{\grsquare}}}(\Delta r)$ (Eq.~\ref{eq:Gaussian}) are fitted to the two data sets (solid lines).
The length at which $C_{\text{I}}^{\text{\grsquare}}$ decays to $1/{\text{e}}$ (indicated for the $x$ direction) is related to the speckle area $A_{\text{S}}^{\text{\grsquare}}$.
}
\label{fig:CI_deltar}
\end{figure}

To quantify the characteristic length scale of the fluctuations, i.e.~the speckle area $A_{\text{S}}^{\text{\grsquare}}$ (Eq.~\ref{eq:AS}), the spatial intensity correlation function $C_{\text{I}}^{\text{\grsquare}}(\Delta {\mathbf{r}})$ (Eq.~\ref{eq:intautocorr}) is determined from the intensity $I_{\text{D}}^{\text{\grsquare}}({\mathbf{r}})$.
It is separately calculated in $x$ and $y$ direction (\cref{fig:CI_deltar}) to account for the slightly elliptical beam (\cref{sec:setup}).
The prediction for a Gaussian beam detected by a square pixel $C_{\text{I}}^{{\text{\grsquare}}}(\Delta \mathbf{r})$ (Eq.~\ref{eq:Gaussian}) is fitted to the experimental data sets.
Despite the approximation of the top-hat beam by a Gaussian beam, it describes the data very well.
The small deviations at large $\Delta \mathbf{r}$ indicate some non-Rayleigh statistics.
This is also suggested by the slightly too large contrast $c$ (\cref{tab:speckles}) and small deviations of the intensity probability distribution $p(I_{\text{D}}^{\text{\grsquare}})$ from the ideal exponential case, and has been observed already previously~\cite{Bromberg2014}.
Furthermore, there are small fluctuations at large $\Delta r$ which are attributed to the circular apertures.
The lengths at which $C_{\text{I}}^{\text{\grsquare}}(\Delta \mathbf{r})$ decays to $1/{\text{e}}$, $\Delta r_x$ and $\Delta r_y$ (\cref{fig:CI_deltar}), provide a measure of the speckle sizes in $x$ and $y$ directions, respectively, and the speckle area $A_{\text{S}}^{\text{\grsquare}} = \pi (\Delta r_x^2{+}\Delta r_y^2)$.
They indicate slightly elliptical speckles with an axial ratio of about $1.1$, consistent with the elliptical beam (Sec.~\ref{sec:setup}).

For an effective measurement area much smaller than the speckle area ($A_{\text{m}}^{\text{\grsquare}} \ll A_{\text{S}}^{\text{\grsquare}}$), hence well above the Nyquist limit $A_{\text{S}}^{\text{\grsquare}} \approx 2A_{\text{m}}^{\text{\grsquare}} = 2\:\text{px}$, equivalent information can be obtained from the width of the power spectral density (Fig.~\ref{fig:IvsPSD}, right), which is inversely proportional to the width of the spatial correlation function~\cite{Kirkpatrick2007,Kirkpatrick2008}.
With decreasing speckle area $A_{\text{S}}^{\text{\grsquare}}$, indeed the peak at low frequencies becomes smaller and broader (\cref{fig:IvsPSD}, top to bottom), consistent with the findings based on $C_{\text{I}}^{{\text{\grsquare}}}(\Delta \mathbf{r})$.

\subsubsection{Time-varying Speckle Pattern}
\label{sec:rotSpec}

\begin{figure} 
\includegraphics[width=0.9\linewidth]{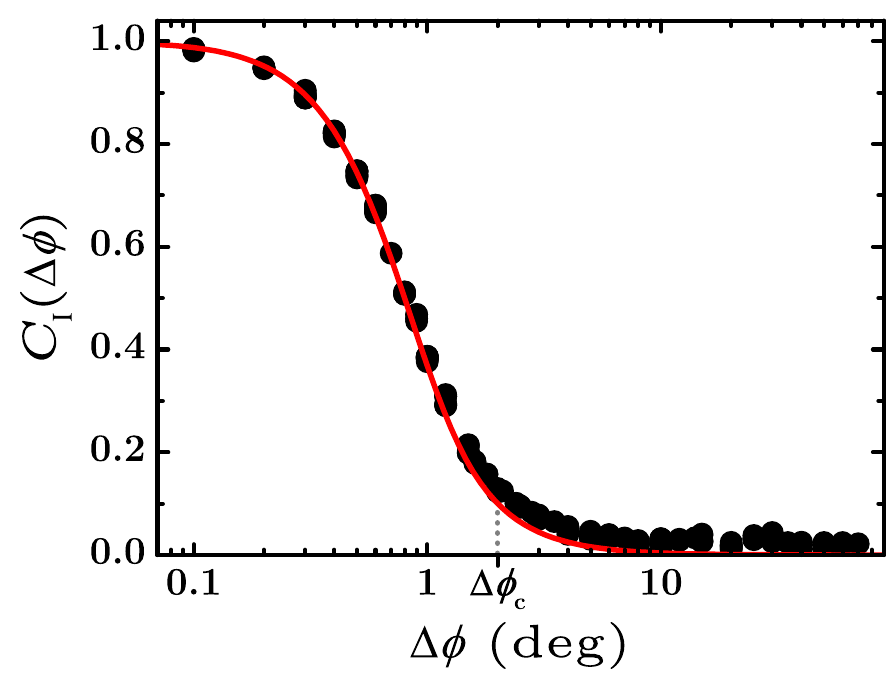}
\caption[Correlation coefficient and intensity distribution for rotating diffuser]
{(Color online) Angular intensity correlation function $C_{\text{I}}^{\text{\grsquare}}(\Delta \phi)$ based on speckle patterns obtained with orientations of the diffuser that differ by $\Delta \phi$ (Eq.~\ref{eq:ImageCorr}).
The red line represents the calculation based on Eq.~\ref{eq:CI_phi_r}.
Experimental conditions BE~5$\times$, which implies a speckle area $A_\text{S}^{\text{\grsquare}} = 5.1\:\upmu$m (Tab.~\ref{tab:speckles}), a square field of view with lateral length $L_\text{v} = 202.2\:\upmu$m and an exposure time of $1.1$ms.
\label{fig:StepCorrelations}
}
\end{figure}

Our set-up offers the possibility to rotate the diffuser around the optical axis.
While a rotation does not change the intensity statistics, the actual speckle pattern is changed and represents another realization, provided the rotation angle $\Delta \phi$ was large enough.
The correlation between two speckle patterns, $I_{\text{D}}^{\text{\grsquare}}(\phi, {\mathbf{r}})$ and $I_{\text{D}}^{\text{\grsquare}}(\phi{+}\Delta \phi, {\mathbf{r}})$ is quantified by the angular correlation function, namely
\begin{equation}
C_{\mathrm{I}}^{\text{\grsquare}}(\Delta \phi) = \frac{\left \langle I_{\text{D}}^{\text{\grsquare}}(\phi, \mathbf{r}) I_{\text{D}}^{\text{\grsquare}}(\phi{+}\Delta \phi,\mathbf{r}) \right \rangle}{\left \langle I_{\text{D}}^{\text{\grsquare}}\right \rangle^2} - 1 \ \text{,}
\label{eq:ImageCorr}
\end{equation}
where $\langle I_{\text{D}}^{\text{\grsquare}} \rangle$ is independent of the angle $\phi$ and $\langle \; \rangle$ represents an average over all pixels, i.e.~all ${\bf r}$, and realizations.
As expected, $C_{\text{I}}^{\text{\grsquare}}(\Delta \phi)$ decreases with increasing $\Delta \phi$ (\cref{fig:StepCorrelations}).
Only very small correlations, say $10\:\%$, are observed beyond $\Delta \phi_{\text{c}} \approx 2^\circ$.
Thus, rotations with $\Delta \phi \gg \Delta \phi_{\text{c}}$ are expected to result in essentially uncorrelated realizations of the speckle pattern.

The definition of $C_{\mathrm{I}}^{\text{\grsquare}}(\Delta \phi)$ is analogous to the spatial intensity correlation function $C_{\mathrm{I}}(\Delta {\bf r})$ (Eq.~\ref{eq:intautocorr}), which can be used to calculate $C_{\mathrm{I}}^{\text{\grsquare}}(\Delta \phi)$.
A rotation of the diffuser by $\Delta \phi$ implies a displacement of the speckle pattern by $\boldsymbol\Delta \boldsymbol\phi \times {\bf r}$, which depends on the distance $r = | {\bf r} |$ from the optical axis around which the speckle pattern is rotated.
Thus, $I_{\text{D}}^{\text{\grsquare}}(\phi, \mathbf{r}) I_{\text{D}}^{\text{\grsquare}}(\phi{+}\Delta \phi,\mathbf{r}) = I_{\text{D}}^{\text{\grsquare}}(\phi, \mathbf{r}) I_{\text{D}}^{\text{\grsquare}}(\phi,\mathbf{r}{-}\boldsymbol\Delta \boldsymbol\phi{\times}{\bf r})$, which relates $C_{\mathrm{I}}^{\text{\grsquare}}(\Delta \phi)$ to $C_{\mathrm{I}}^{\text{\grsquare}}(\Delta {\bf r})$. Averaging over a circular field of view with radius $R_{\text{v}}$ and square pixels, and using the correlation function for a Gaussian beam detected by square pixels, $C_{\text{I}}^{\text{\grsquare}}(\Delta r)$ (Eq.~\ref{eq:Gaussian}), yields
\begin{align}
C_{\mathrm{I}}^{{\text{\grsquare}}}(\Delta \phi)
   =& \frac{1}{\pi R_{\text{v}}^2} \int_0^{R_{\text{v}}}{C_{\text{I}}^{{\text{\grsquare}}}(r \Delta \phi) \; 2 \pi r \; {\mathrm{d}}r} \nonumber \\
   =& \frac{A_{\text{s}}}{\pi R_{\text{v}}^2} \; \frac{1}{\Delta \phi^2} \left \{ 1 - \exp{\left (-\frac{\pi R_{\text{v}}^2}{A_{\text{s}}} \Delta \phi^2 \right )} \right \}  \; .
\label{eq:CI_phi_r}
\end{align}
For a square field of view with size $L_\text{v}^2$ and square pixels, the corresponding relation involves the error function. However, it can be approximated by a circular field of view, i.e.~\cref{eq:CI_phi_r}, with an effective radius $R_\text{v} \approx 0.57 \, L_\text{v}$, which corresponds to a slightly larger effective area. This prediction is confirmed by the experimental data~(\cref{fig:StepCorrelations}).

To fully characterize time-varying speckle patterns, the angular velocity has to be considered.
This is similar to the situation in speckle contrast analysis, imaging applications and light scattering~\cite{Briers1996, Duncan2008, Scheffold2012, Briers2013}.
If the diffuser is rotated, and hence the speckle pattern changed, faster than the particles can follow, i.e.~than their relaxation time, the colloidal particles effectively experience a temporally averaged and hence microscopically flat intensity pattern instead of a speckle pattern (\cref{fig:profile}).
Then, only time-averaged intensities are of interest.
Both, averages over many images with short exposure times as well as individual images with long exposure times, are considered.
With appropriate camera parameters, both procedures yield equivalent time-averaged intensities~\cite{Scheffold2012}.
The average over many realizations indeed shows significantly reduced fluctuations compared to the static speckle pattern (\cref{fig:profile}A,B).
Nevertheless, a small modulation remains, even in the azimuthal average (\cref{fig:profile}C).


\subsection{Random Potential Energy Landscape}
\label{sec:rPEL}

Having investigated the speckle patterns, we now consider their effect on spherical colloidal particles that are characterized by the weight function $D^{\odot}(\mathbf{r})$ (Eq.~\ref{eq:D_particle}, \cref{sec:potential}).
The effect of a speckle pattern can be described by an external potential $U({\bf r})$, the rPEL (as the one shown in Fig.~\ref{fig:IplusUplusConv}C).
We will now determine the properties of $U({\bf r})$.

\subsubsection{Time-averaged local particle density}

\begin{figure} 
\includegraphics[width=0.8\linewidth]{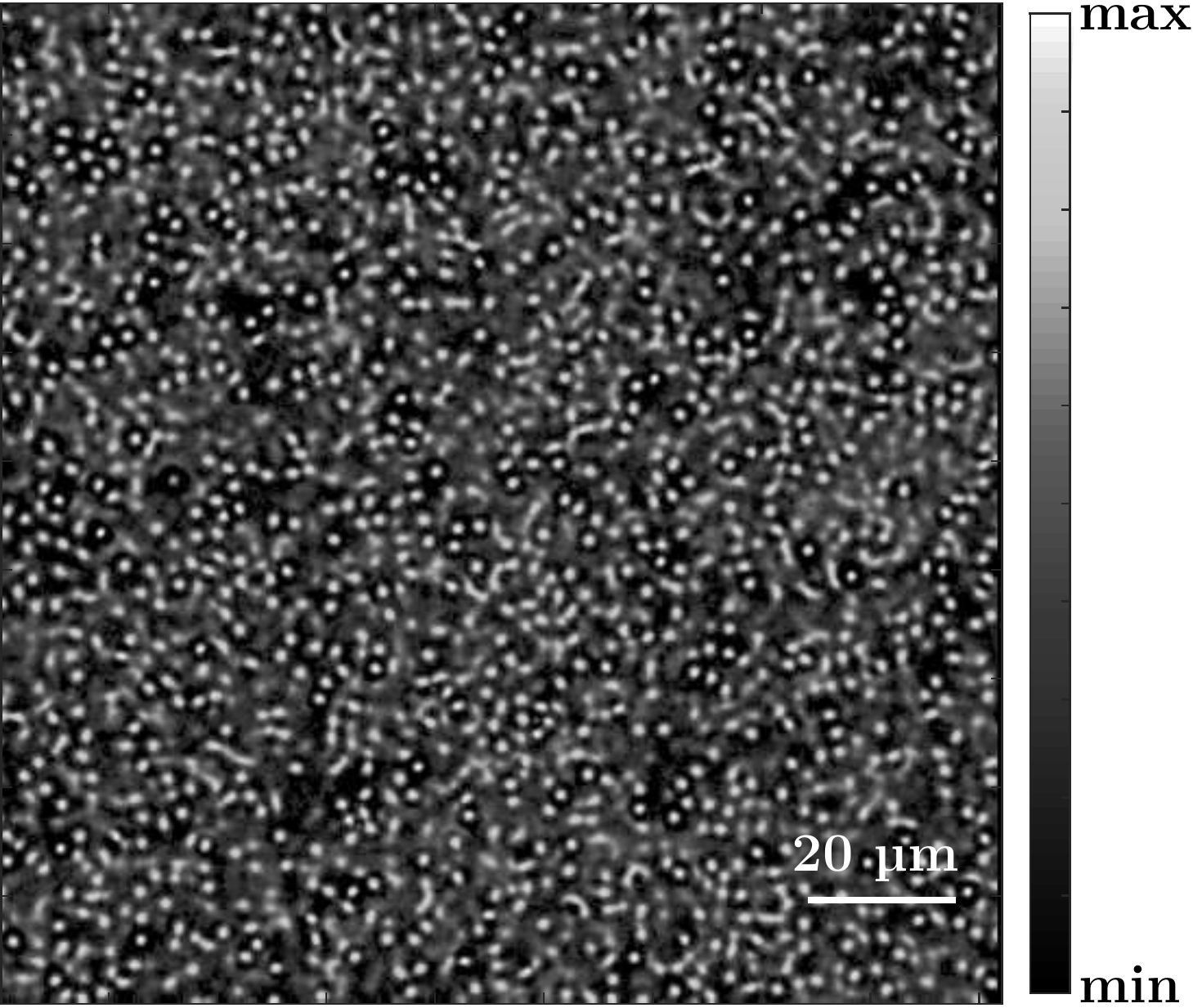}
\caption[$\langle\rho(\mathbf{r})\rangle$]
{Time-averaged local particle density $\rho({\mathbf{r}})$ of a (quasi) two-dimensional layer of spherical polystyrene particles with sulfonated chain ends with radius $R=1.4\; \upmu$m, polydispersity 3.2~\% and mean surface density $\langle \rho \rangle = 0.25$ in a speckle pattern (BE 5$\times$, Tab.~\ref{tab:speckles}) created using a moderate laser power ($P_{\text{L}} = 1640$~mW).
About $37\,000$~images at $3.75$~fps (AVT, Pike F032B) were recorded and averaged.
Densities are represented as grey levels (logarithmic scale in arbitrary units).
}
\label{fig:rho_xy}
\end{figure}

The speckle pattern affects the distribution of particles.
It is quantified by the time-averaged local particle density $\rho({\mathbf{r}})$ which is determined from the particle locations~\cite{Crocker1996}.
The density $\rho({\mathbf{r}})$ for a (quasi) two-dimensional layer of particles with a mean surface fraction $\langle \rho \rangle = 0.25$, i.e.~about $1200$ particles in a field of view of $171\times171~\upmu\text{m}^2$, is shown in figure~\ref{fig:rho_xy}.
A qualitative inspection reveals that $\rho({\mathbf{r}})$ resembles some of the characteristics of the rPEL $U({\mathbf r})$ (\cref{fig:IplusUplusConv}C).
It exhibits random fluctuations with a comparable characteristic length scale, but also longer-ranged correlations.
Furthermore, the maxima of $\rho({\mathbf{r}})$ are more pronounced while the saddle points and minima are blurred.
Within reasonable measurement times, the low $\langle \rho \rangle$ and the strongly disordered potential hence do not provide sufficient statistics to obtain space-resolved information on $\rho({\mathbf{r}})$ and thus the potential $U({\mathbf{r}})$.
This suggests to investigate samples with larger $\langle \rho \rangle$.
However, a straight-forward determination of $U({\mathbf{r}})$ from $\rho({\mathbf{r}})$ through the Boltzmann distribution requires that particle--particle interactions can be neglected and thus that the sample is dilute.
In more concentrated systems, the determination of $U({\mathbf{r}})$ requires to apply more involved methods, e.g., liquid-state theory~\cite{Sengupta2005} or Inverse Monte Carlo Simulations~\cite{Bahukudumbi2007}.
This is beyond the scope of the present work.

\subsubsection{Convolution with the Weight Function of a Spherical Particle}

To avoid this complication, we investigate the convolution of the speckle pattern with the weight function of a spherical particle, $D^{\odot}(\mathbf{r})$, and, instead of the full $U({\bf r})$, determine the statistics of $U({\bf r})$, namely the distribution of its values, the magnitude of its fluctuations and its correlation area.
In the case of a particle exposed to a light field, $D^{\odot}(\mathbf{r})$ describes the susceptibility of the particle to light (Eq.~\ref{eq:D_particle}), but is formally identical to a detector efficiency.
The convolution of $D^{\odot}(\mathbf{r})$ with the intensity pattern $I({\bf r})$ yields the total intensity $I_{\text{D}}^\odot({\mathbf{r}})$ that is `detected' by a particle at position ${\bf r}$ (Eq.~\ref{eq:ID}).
Due to the light--matter interaction~\cite{Ashkin1986, Ashkin1992,Pelton2004,Bohren2004,Rohrbach2005,Bonessi2007}, $I_{\text{D}}^\odot({\mathbf{r}})$ represents an external potential $U({\mathbf{r}})=I_{\text{D}}^\odot({\mathbf{r}})$ imposed on the particle, that is the rPEL.
Since $D^{\odot}(r)$ takes into account the volume of the particle, the extended colloidal particle at position $\mathbf{r}$ in the speckle pattern $I({\mathbf{r}})$ can be regarded as a point-like particle in the potential $U({\bf r})=I_{\text{D}}^\odot({\mathbf{r}})$.
This procedure and a typical $U({\mathbf{r}})$ are illustrated in \cref{fig:IplusUplusConv}.
It has already successfully been applied to micron-sized colloidal particles in a one-dimensional rPEL; experiments and simulations yielded consistent results~\cite{Hanes2012a}.

Potentials $U({\mathbf{r}})$ obtained by convolving experimental speckle patterns $I_{\text{D}}^{\text{\grsquare}}({\mathbf{r}})$ with the weight function $D^{\odot}(\mathbf{r})$ (Eqs.~\ref{eq:ID}, \ref{eq:D_particle})~\cite{Loudiyi1992, Pelton2004, Jenkins2008, Hanes2012a} are quantitatively investigated in the following.
This allows us to test whether $p(I_{\text{D}}^\odot)=p(U)$ can be described by a Gamma distribution (Eq.~\ref{eq:PIP}) and to find an approximation for the parameter $M$.

\begin{figure*} 
\includegraphics[width=0.80\linewidth]{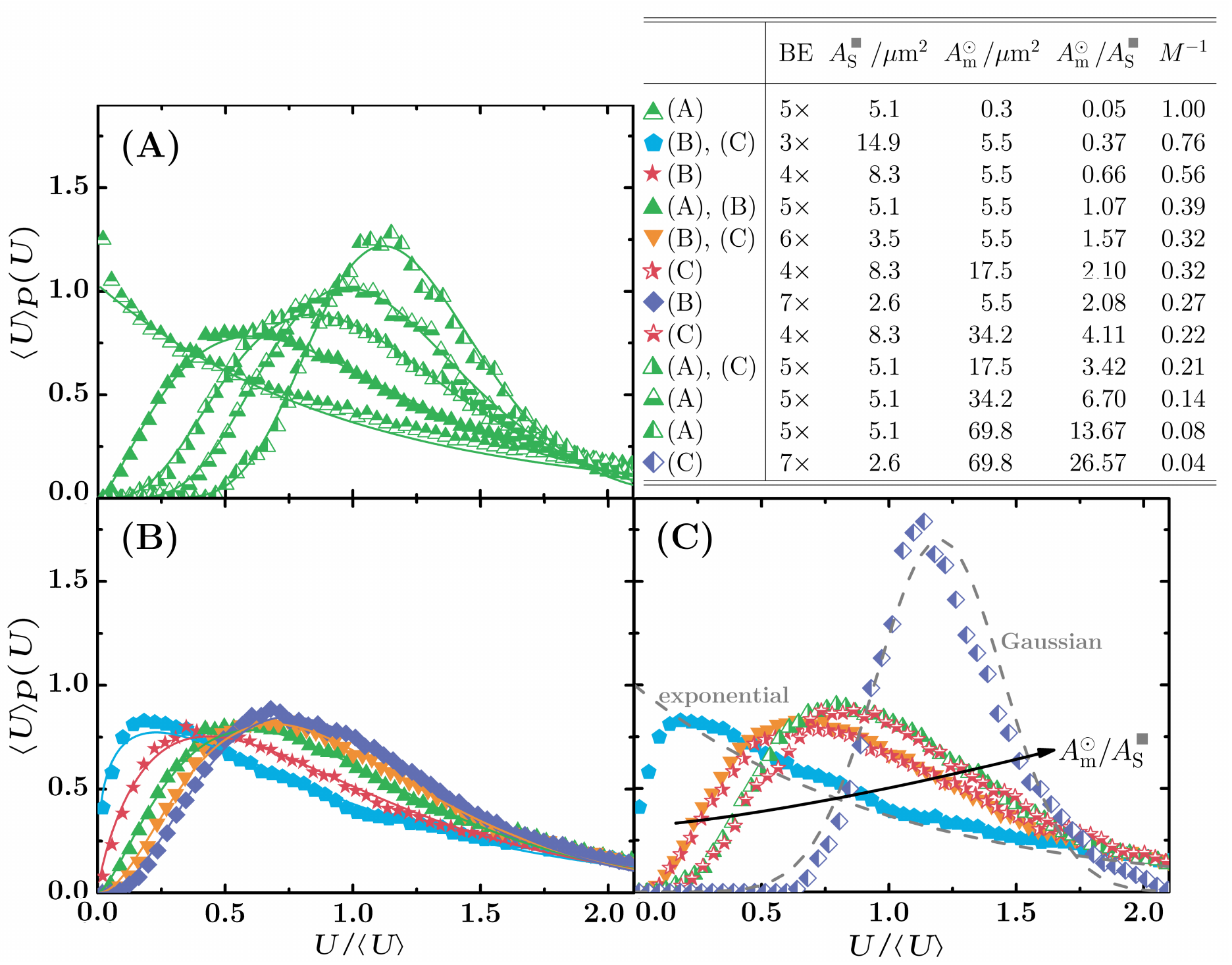}
\caption[Distribution of potential values $p(U)$ calculated by convolution in dependency of the particle radius $R$]
{Normalized distribution of potential values $\langle U \rangle \, p(U)$ calculated by the convolution of an experimentally-determined speckle pattern $I_{\text{D}}^{\text{\grsquare}}(\mathbf{r})$ with the weight function $D^{\odot}(\mathbf{r})$ of a colloidal particle for
(A) different effective particle areas $A_{\text{m}}^{\odot}$ and constant speckle area $A_{\text{S}}^{\text{\grsquare}} = 5.1~\upmu\text{m}^2$ (BE~5$\times$, Tab.~\ref{tab:speckles}),
(B) different $A_{\text{S}}^{\text{\grsquare}}$ and constant $A_{\text{m}}^{\odot} = 5.5~\upmu\text{m}^2$ and
(C) different $A_{\text{m}}^{\odot}/A_{\text{S}}^{\text{\grsquare}}$.
The different conditions and the inverse of the fit parameter $M$ are summarized in the table.
The solid lines represent fits by Gamma distributions (Eq.~\ref{eq:PIP}) and dashed lines in (C) fits by an exponential and a Gaussian distribution as indicated.
}
\label{fig:PURconv}
\end{figure*}

On a qualitative level, $U({\mathbf{r}})$ appears washed out compared to the speckle pattern (\cref{fig:IplusUplusConv}) due to the convolution with $D^{\odot}(\mathbf{r})$.
The magnitude of the fluctuations is reduced and their characteristic length scale is increased, in particular if the particle is large.
This is consistent with experimental observations (Fig.~\ref{fig:rho_xy}).

The distribution of potential values $p(U)$ depends on the speckle area $A_{\text{S}}^{\text{\grsquare}}$ (Eq.~\ref{eq:AS}) and the effective particle area $A_{\text{m}}^{\odot} = (8\pi/9) R^2$ (Eq.~\ref{eq:Aeffcirc}).
If $A_{\text{S}}^{\text{\grsquare}}$ is kept constant, the effect of the particle area $A_{\text{m}}^{\odot}$ on $p(U)$ can be studied (\cref{fig:PURconv}A).
We consider particles with radii in the range $0.3~\upmu$m~$\le R \le 5.0~\upmu$m, which are large enough to be observed with the microscope.
A small $R$ or $A_{\text{m}}^{\odot}$ leads to an almost exponential distribution and develops into an approximately Gaussian distribution as $A_{\text{m}}^{\odot}$ increases.
Correspondingly, for constant $A_{\text{m}}^{\odot}$ but decreasing $A_{\text{S}}^{\text{\grsquare}}$, a similar transition from an exponential to an approximately Gaussian distribution is observed (Fig.~\ref{fig:PURconv}B).
More general, similar distributions are obtained for comparable $A_{\text{m}}^{\odot}/A_{\text{S}}^{\text{\grsquare}}$  (Fig.~\ref{fig:PURconv}C) and thus $p(U)$ appears to only depend on this ratio with its shape changing from an almost exponential distribution to an approximately Gaussian distribution upon increasing $A_{\text{m}}^{\odot}/A_{\text{S}}^{\text{\grsquare}}$.

For all $A_{\text{m}}^{\odot}$ and $A_{\text{S}}^{\text{\grsquare}}$, a Gamma distribution (Eq.~\ref{eq:PIP}) is fitted to the data.
The Gamma distribution describes the distribution of potential values $p(U)$ well and only depends on $A_{\text{m}}^{\odot}/A_{\text{S}}^{\text{\grsquare}}$.
Only small deviations are observed, similar to those reported before~\cite{Ducharme2007}.
They are attributed to the approximations leading to the Gamma distribution~\cite{Goodman2007}, e.g.~a Gaussian instead of a top-hat beam and the presence of finite optical components and detector pixels, and a possible effect of the (very few) particles on the speckle pattern (\cref{sec:speckles}).

\begin{figure} 
\includegraphics[width=0.93\linewidth]{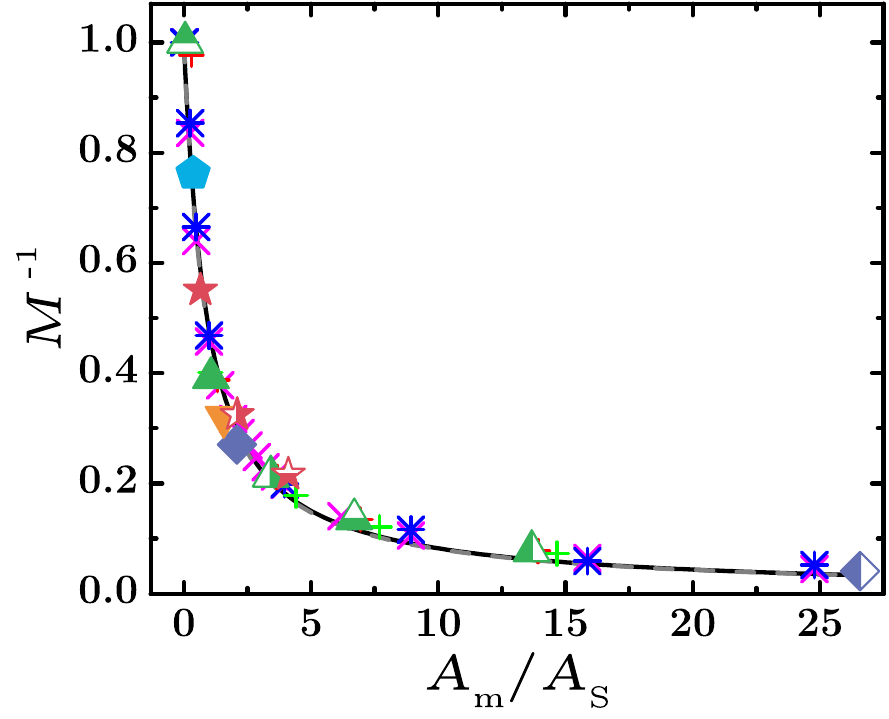}
\caption[$M^{-1}(A_{\text{eff}}/S)$]
{(Color online) Parameter $M^{-1}$, which quantifies the fluctuations of the potential $U({\bf r})$, as a function of the ratio of the effective particle area $A_{\text{m}}$ and the speckle area $A_{\text{S}}$, for different conditions (as indicated in~\cref{fig:PURconv}) as well as 
(magenta~$\times$) cubic (instead of spherical) particles with different sizes, i.e. effective particle areas $A_{\text{m}}^\boxdot$, in the experimental condition BE~5$\times$, 
(red +, green +) spherical particles in an intensity pattern which has been smoothed over $5 \times 5$ and $10 \times 10$ pixels, respectively, and
(blue asterisk) a point-like particle, i.e.~no convolution, in an intensity pattern which has been binned over different numbers of pixels ($1\times1~\text{to}~50\times50$).
The dashed grey and  solid black lines represent predictions for a Gaussian beam and a square (Eq.~\ref{eq:M2}) and circular (Eq.~\ref{eq:Mcirc}) detector, respectively.
}
\label{fig:1overM}
\end{figure}

The fit of the Gamma distribution to the data yields the parameter $M$ (Eq.~\ref{eq:PIP}, Fig.~\ref{fig:PURconv}), which is related to the contrast $c$ and standard deviation $\sigma$.
In the case of the potential, $\sigma$ represents the magnitude of the fluctuations or `roughness' of the random potential $U({\bf r})$.
Thus we consider $M^{-1} \sim \sigma^2$.
Independent of the specific particle and speckle sizes, $M^{-1}$ only depends on $A_{\text{m}}^{\odot}/A_{\text{S}}^{\text{\grsquare}}$ and decreases with increasing $A_{\text{m}}^{\odot}/A_{\text{S}}^{\text{\grsquare}}$ (\cref{fig:1overM}).
Thus, the magnitude of the fluctuations only depends on the number of speckles that interact with a particle.

\begin{figure} 
\includegraphics[width=0.96\linewidth]{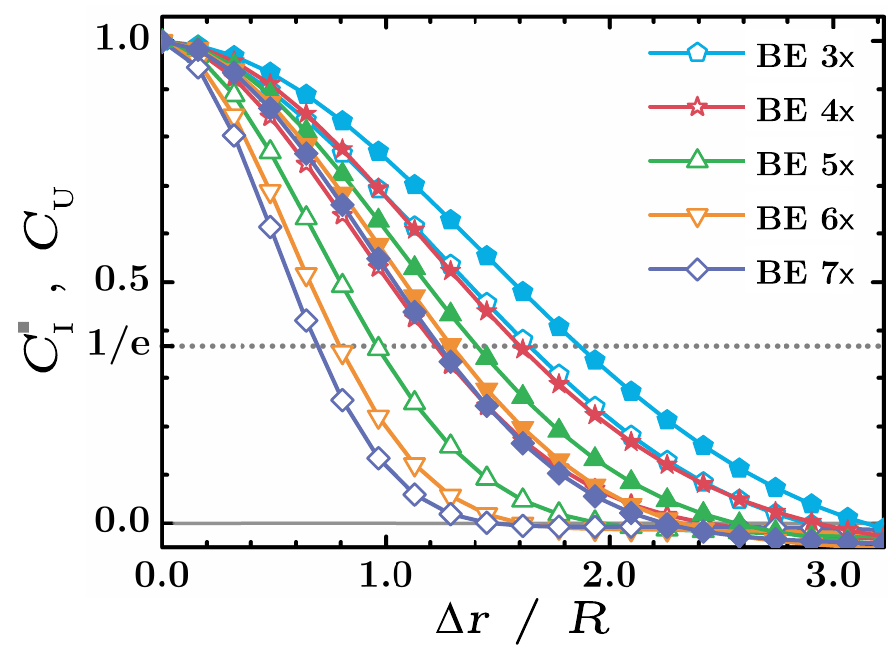}
\caption[$C_U$ and $p(U)$]
{(Color online) Intensity, $C_{\text{I}}^{\text{\grsquare}}(\Delta r)$ (connected open symbols), and potential, $C_{\text{U}}(\Delta r)$ (connected filled symbols), correlation functions as a function of $\Delta r$, which are based on the intensity $I_{\text{D}}^{\text{\grsquare}}({\mathbf{r}})$ and its convolution with the weight function $D^{\odot}(\mathbf{r})$ of a particle with radius $R=1.4\,\upmu$m, respectively~(\cref{tab:speckles}).
The lengths at which the correlation functions decay to $1/{\text{e}}$ are related to the speckle areas $A_{\text{S}}^{\text{\grsquare}}$ and the correlation areas of the potential, $A_{\text{S}}^{\text{U}}$, respectively, and hence to the effective correlation area of the weight function $A_{\text{S}}^{\odot}$.
}
\label{fig:CIvsCU}
\end{figure}

The correlation area $A_{\text{S}}^{\text{U}}$ of the potential $U({\bf r})$ is obtained from the length at which the correlation function $C_U(\Delta r)$ decays to $1/{\text{e}}$ (Fig.~\ref{fig:CIvsCU}, Tab.~\ref{tab:speckles}).
The correlation area of the intensity or speckle area $A_{\text{S}}^{\text{\grsquare}}$ decreases with increasing beam size and hence also $A_{\text{S}}^{\text{U}}$.
The difference between both values is the effective correlation area of the weight function; $A_{\text{S}}^{\odot} = A_{\text{S}}^{U} - A_{\text{S}}^{\text{\grsquare}}$ (Tab.~\ref{tab:speckles}).
The value of $A_{\text{S}}^{\odot}$ only depends on the weight function $D^\odot({\bf r})$ as long as the effective particle area $A_{\text{m}}^{\odot}$ is larger than the speckle area $A_{\text{S}}^{\text{\grsquare}}$, consistent with the data (Tab.~\ref{tab:speckles}).

\subsubsection{Convolution with Other Weight Functions}

Instead of convolving the intensity $I_{\text{D}}^{\text{\grsquare}}({\mathbf{r}})$ with the weight function of spherical particles, $D^{\odot}(\mathbf{r})$, it is convolved with the weight function of cubic particles, $D^{\boxdot}({\mathbf{r}})$, with different effective particle areas $A_{\text{m}}^\boxdot$.
A similar $M^{-1}(A_{\text{m}}^{\boxdot}/A_{\text{S}}^{\text{\grsquare}})$ is obtained (\cref{fig:1overM}, magenta~$\times$).
This indicates that the precise shape of the particle is not crucial, as long as it has the same effective particle area $A_{\text{m}}$.

Furthermore, the effect of smoothing is investigated.
The intensity $I_{\text{D}}^{\text{\grsquare}}({\mathbf{r}})$ is smoothed using a filter before it is convolved with $D^{\odot}(\mathbf{r})$.
The filter replaces each pixel's intensity $I_{\text{D}}^{\text{\grsquare}}({\mathbf{r}})$ by the average intensity of the $n \times n$ pixels surrounding the pixel, $I_{\text{D}}^{\text{\makebox(6,4)[lb]{\textcolor{black}{\FilledSquareShadowC}}}}({\mathbf{r}})$.
Smoothing is equivalent to the convolution of the intensity with $D^{\boxdot}({\mathbf{r}})$ described above, but $I_{\text{D}}^{\text{\textcolor{black}{\FilledSquareShadowC}}}({\mathbf{r}})$ in addition is subsequently convolved with $D^{\odot}({\mathbf{r}})$.
Both yield virtually identical results (\cref{fig:1overM}, red +, green +) 
, as long as the smoothing is taken into account in the calculation of $A_{\text{S}}^{U}$, i.e.~$A_{\text{S}}^{U} = A_{\text{S}}^{\odot} + A_{\text{S}}^{\text{\textcolor{black}{\FilledSquareShadowC}}} + A_{\text{S}}^{\text{\grsquare}}$.
This becomes increasingly more significant as smoothing extends over larger areas.

Finally, $I_{\text{D}}^{\text{\grsquare}}({\mathbf{r}})$ is binned into larger `meta pixels' resulting in a larger effective measurement area $A_{\text{m}}^\boxplus$, but smaller number of (meta) pixels.
This is in contrast to smoothing, where the number of pixels is maintained.
The corresponding intensity $I_{\text{D}}^\boxplus({\mathbf{r}})$ mimicks a camera with larger but less pixels.
Hence, the number of speckles in the effective measurement area is increased, $A_{\text{m}}^\boxplus / A_{\text{S}}^{\text{\grsquare}} > A_{\text{m}}^{\text{\grsquare}} / A_{\text{S}}^{\text{\grsquare}}$.
Nevertheless, for a sufficient number of meta pixels (above about $20$), $p(I_{\text{D}}^\boxplus)$ can be described in good approximation by a Gamma distribution (Eq.~\ref{eq:PIP}, data not shown) and $M^{-1}(A_{\text{m}}^\boxplus/A_s^{\text{\grsquare}})$ shows the same dependence on $A_{\text{m}}^\boxplus/A_{\text{S}}^{\text{\grsquare}}$ (\cref{fig:1overM}, blue asterisk). 

These findings suggest that the dependence of $M^{-1}$ on $A_{\text{m}}/A_{\text{S}}$ does not strongly depend on the experimental conditions as long as they are properly taken into account through $A_{\text{m}}$ and $A_{\text{S}}$.
Thus, our experimental situation, namely a top-hat beam and a particle as `detector', appears  well approximated by a Gaussian beam and a square or circular detector.
Indeed, $M$ as given by Eqs.~\ref{eq:M2} or \ref{eq:Mcirc}, which both only depend on the ratio $A_{\text{m}}/A_{\text{S}}$, reproduces our findings very well (\cref{fig:1overM}, lines).
This confirms previous experimental results for similar, but not identical, speckle patterns and optical geometries~\cite{Skipetrov2010, Li2012}.
We hence established an appropriate description of the statistics of the rPEL, $U({\mathbf{r}})$, imposed on the colloidal particles.
In particular, the distribution of potential values can be characterized by a Gamma distribution (Eq.~\ref{eq:PIP}) and the parameter $M^{-1}$, quantifying the magnitude of its fluctuations, by Eq.~\ref{eq:M2} or \ref{eq:Mcirc}.


\section{Conclusions}
\label{sec:conclusions}

We experimentally realize random potential energy landscapes exploiting the interaction of matter with light.
Colloidal particles are investigated which act as `detectors' in a random intensity pattern, that is laser speckles.
The speckle pattern is produced using an optical set-up which is based on a special diffuser.
The diffuser creates a top-hat beam containing a speckle pattern.
This speckle pattern is quantitatively characterized.
In the standard experimental conditions, the intensity distribution is found to follow an exponential distribution with the normalized standard deviation or contrast being close to one, which indicates that fully developed speckles are formed.
Their size can be controlled through the size of the illuminating laser beam.

The interaction of the particle with the speckle pattern is described analogous to a detector recording the intensity.
However, the intensity that is `detected' by the particle represents an external potential that is imposed on the particle, the rPEL. 
It is found that the distribution of energy values of the rPEL can be described by a Gamma distribution and approximations for the standard deviation of the distribution are identified.
Using these approximations, thus, the statistics of the rPEL can quantitatively be described.
These relations together with the set-up, can be exploited to produce rPELs with the desired distribution of energy values and correlation lengths, where the shape of the distribution can be varied in a broad range, from exponential to Gaussian.

When colloidal particles are exposed to such an intensity pattern, that is an rPEL, their spatial arrangement and dynamics will be affected as demonstrated previously~\cite{Hanes2012a, Hanes2013,Evers2013a,Evers2013b} and in agreement with theoretical predictions~\cite{Bouchaud1990, Dean2007, Sengupta2005, Isichenko1992, Goychuk2014, Banerjee2014, Wales2004, Zwanzig1988}.
In these previous studies, the speckle patterns have been created using a spatial light modulator~\cite{Hanes2009}.
Compared to this method, the present set-up offers a much larger field of view and thus the possibility to simultaneously observe a much larger number of particles.
The distribution of potential energy values and their spatial correlation furthermore are tunable.
In addition, the diffuser can be rotated and hence the speckle pattern varied.
If this is faster than the particle dynamics, the particles experience a time-averaged and hence flat effective potential.
Radiation pressure still pushes them towards the wall and the increased hydrodynamic interactions slow them down.
Therefore, the effect of hydrodynamic wall--particle interactions can be determined independently.


\section*{Acknowledgments}

We thank R.~Capellmann, M.~Escobedo-Sanchez, S.~Gl\"ockner, F.~Platten, D.~Wagner and C.~Zunke for very helpful discussions and suggestions, and the Deutsche Forschungsgemeinschaft (DFG) for financial support within the SFB-TR6.


\bibliography{RPEL_arxive}

\end{document}